\documentclass[aps,showpacs,preprintnumbers,amsmath,amssymb,twocolumn,tightenlines
]{revtex4}

\usepackage{bm}
\usepackage{amsmath}
\usepackage{graphicx}
\usepackage{epic}
\usepackage{eepic}

\newcommand{\F}{\mathcal{F}}
\newcommand{\e}{\varepsilon}

\newcommand{\ee}{$e^+e^-~$}
\newcommand{\mm}{$\mu^+\mu^-~$}

\begin{document}

    \bibliographystyle{apsrev}

    \title {Production of high energy particles in laser and Coulomb fields and \ee antenna}
               
    \author{M. Yu. Kuchiev}
                \email[email:]{kmy@phys.unsw.edu.au}
    \affiliation{School of Physics, University of New South Wales,
      Sydney 2052, Australia}
    \date{\today}

\begin{abstract}

A strong laser field and the Coulomb field of a nucleus can produce \ee pairs. It is shown for the first time that there is a large probability that electrons and positrons created in this process collide after one or several oscillations of the laser field. These collisions can take place at high energy resulting in several phenomena. The quasielastic collision \ee $\rightarrow$ \ee allows acceleration of leptons in the laser field to higher energies. The inelastic collisions allow production of high energy photons $e^+e^-\rightarrow 2\gamma$ and muons $e^+e^-\rightarrow \mu^+\mu^-$. 
The yield of high-energy photons and muons produced via this  mechanism exceeds  exponentially their production through conventional direct creation in laser and Coulomb fields. A relation of the phenomena considered with the antenna-mechanism of multiphoton absorption in atoms is discussed.
\end{abstract}


    \maketitle

Creation of \ee pairs in a laser field has similarity with the Schwinger mechanism of pair production  by a static electric field \cite{Sch51}. The \ee production by laser and Coulomb fields may also be compared with Ref.\cite{BH34}, being sometimes called the nonlinear Bethe-Heitler process. This mechanism of \ee production is considered in literature as a possible candidate for experimental studies for a new generation of lasers \cite{Rin01}. Various theoretical aspects of this process were discussed in Refs. \cite{Yak66,Mit87,MVG03, MVG032, MVG04,DP98,AAMS-03,MMHJK06,KR07}. 
A popular approach to the problem uses Volkov wave functions for leptons in a laser field \cite{Vol35}, which is close to the Keldysh approximation \cite{Kel65} for multiphoton phenomena in atoms \cite{NR66,PPT66,Fai73,Rei80,GK97}. The recent Ref. \cite{KR07} simplified the formalism for the \ee production and derived several important characteristics of the problem in simple analytical form.

The leptons $e^+$ and $e^-$ initially  created in the vicinity of the nucleus, propagate out of the Coulomb center along classical trajectories, which exhibit wiggling in the laser field. Obviously,  $e^+$ and $e^-$  follow different trajectories. However, we will see that the wiggling can force the leptons to collide after one or several periods of laser oscillations. 
The probability of this collision proves surprisingly large. At the moment of collision the energy of the \ee pair can be high, which results in several interesting phenomena. The quasielastic collision \ee $\rightarrow$ \ee can make energy of a lepton higher than its initial  energy. There is also a possibility for inelastic collisions, which result in production of high-energy particles. 
(The term high-energy in this work is applied energies above the electron mass.)
This includes production of high-energy photons via annihilation \ee $\rightarrow 2\gamma$, and production of heavy particles, such as muons, \ee$\rightarrow \mu^+\mu^-$. 

In the phenomena considered the \ee pair fulfills two major functions. First, it  accumulates energy from the laser field, and then it transfers this energy to other high-energy particles. This reminds closely the work of conventional aerials in radio devices. 
Moreover, physical reasons, which ensure effective accumulation of energy by the \ee pair and by aerials  are very similar. In both cases it is vital that leptons propagate  large distances while exhibiting oscillations in an external electromagnetic field.
This similarity makes it convenient to think about the \ee pair in the phenomena discussed as if it was a particular \ee antenna. A similar idea  was put forward in relation to multiphoton processes in atoms \cite{kuchiev-87,Corkum-93,Schafer-93}, where it was called atomic antenna, or rescattering mechanism. Its validity was confirmed experimentally, see e.g. \cite{Weckenbrock-etal-03,Pedregosa-Gutierrez-etal-04} and references therein.

Consider a linearly polarized laser field, which propagates along the $z$-axis with electric and magnetic fields stretched along the $x$ and $y$ axes.
Describe this field by the vector-potential $e \mathbf{A}(\varphi)= - m \,\xi\, (\sin\varphi,0,0)$,
where $\varphi=\omega(t-z)$, $\omega$ is the laser frequency,
$\xi=eE/m\omega$ is an adiabatic parameter, $e>0$ and $m$ are the absolute value of the electron charge and electron mass ($\hbar=c$). 
Consider the tunneling regime presuming that the laser field is strong, though below the critical value $E_\mathrm{c} =m^2/e$, and $\omega$ is low
\begin{equation}
	\label{tun}
	\F \equiv E/E_\mathrm{c}=eE/m^2\ll 1, \quad\quad \xi=eE/m\omega\gg 1~.
\end{equation}
We will base our discussion on the differential rate of the \ee production $dW_{e^+e^-}$ found in \cite{KR07} 
\begin{align}
	\label{dW}
	dW_{e^+e^-}=&B \exp \bigg(-\frac{ 2\sqrt{3} } {\F} \, \,\frac{\cal{A}}{m^2} \,\bigg) \,\,\frac{ d^3 p_+\, d^3 p_-}{(2\pi)^6}~,
\\	\label{A}
	\cal{A} = &
	\frac{(k_+ - k_-)^2}{8\xi^2}+
	\frac{(k_+ + k_-)^2}{4}
	+\frac{\kappa_+^2+\kappa_-^2}{2}
	\\ \nonumber
 &\quad \quad +	\frac{ 7 (\delta \eta_+^2+\delta \eta_-^2)+ 2\delta \eta_{+} \delta \eta_{-} }{12}~.
\end{align}
Here the states of the positron and electron are described by momenta, which in the Cartesian coordinates read $\mathbf{p}_\pm=(k_\pm,\kappa_\pm,q_\pm)$.
The subscripts $\pm$ mark variables of the positron and electron. 
It is useful to express the longitudinal components of momenta $q_\pm=(m^2+k_\pm^2+\kappa_\pm^2-\eta_\pm^2)/(2\eta_\pm)$ via 
the variables $\eta_\pm$. The $\eta_\pm$ variables
have mean values $\bar {\eta} =m/\sqrt{2}$ \cite{KR07}, deviations from them are called $\delta \eta_\pm= \eta_\pm-\bar {\eta}$. 
The convenience of  $\eta_\pm$ prompts one to use the definition $d^3 p_\pm\equiv dk_\pm 
d\kappa_\pm d\eta_\pm$ for integrations over momenta.
The constant $B$ in Eq.(\ref{dW}) was calculated in \cite{KR07}. Integrating Eq.(\ref{dW}) with this constant one reproduces the rate of the \ee production
\begin{equation}
\label{W}
	W_{e^+e^-}=\frac{ Z^2\alpha^2 m }{ \sqrt{2}\,\,\pi  }\bigg(  \frac{ \F }{ 2\sqrt{3} }\bigg)^3 \exp
	\bigg(  -\frac{ 2\sqrt{3} }{ \F } \,\bigg)~.
\end{equation}
Here $Z$ is the charge of the nucleus, $\alpha=e^2\simeq1/137$.  Eq.(\ref{W}) complies with the results obtained previously (see \cite{KR07,MMHJK06} for discussion and references), which confirms the validity of Eq.(\ref{dW}).

Eqs.(\ref{dW}),(\ref{A}) show that the momenta distribution of the \ee pair has a maximum when
$\mathbf{p}_+=\mathbf{p}_-=\big(0,0,m/(2\sqrt2)\,\big)$. 	
Deviations of momenta from these values are small, proportional to $\sqrt{\F}\,m \ll m$. An important exception is the case when momenta of both leptons satisfy
\begin{equation}
\label{soft}
k_++k_-=0~, \quad \kappa_\pm=\delta \eta_\pm=0~.  	
\end{equation}
Then only the first term in Eq.(\ref{A}) is present. A large factor $\xi^2$ in its denominator makes possible relatively large variations of momenta $|\delta(k_+-k_-)|=2|\delta k_-|$,  $|\delta k_-|\sim \xi  \sqrt{\F}\,m\gg \sqrt{\F}\,m$. 
Consequently, we find that the integration in Eq.(\ref{dW}) is saturated in the close vicinity of the region, where momenta satisfy Eq.(\ref{soft}). We derive a simple and important conclusion that leptons are predominantly created with momenta which satisfy Eq.(\ref{soft}).

An \ee pair is initially created in the close vicinity of the nucleus $r\le m^{-1}$ \cite{KR07}. However, after a short period of time the leptons $e^+$ and $e^-$ go away from the Coulomb center into the region where the laser field is stronger than the Coulomb one. Further propagation of leptons is guided mostly by the laser field. The semiclassical nature of this field allows one to base analysis on the classical trajectories of leptons $\mathbf{r}=\mathbf{r}(\varphi)$, which can be written in a simple form (compare e.g. \cite{LL02-87-sh})
\begin{equation}
	\label{r}
\mathbf{r}(\varphi)=
\frac{1} {\omega \eta } \int_0^ \varphi \mathbf{P}(\varphi')\,d\varphi'~.
\end{equation}
Here  $\mathbf{P}(\varphi)=(K(\varphi), \kappa, Q(\varphi))$ is the instant momentum, 
\begin{align}
\label{PT}
&K(\varphi)=k\mp eA(\varphi)~,
\\\label{PL}
&Q(\varphi)=( m^2+K^2(\varphi)+\kappa^2-\eta^2 )/(2\eta)~.
\end{align}
The instant energy of a lepton is $\varepsilon(\varphi)=Q(\varphi)+\eta$.
In Eqs.(\ref{r})-(\ref{PL}) the subscripts $\pm$  are suppressed, the upper (lower) sign in Eq.(\ref{PT}) corresponds to the case of positron (electron). 
One can anticipate that the \ee pair is created at the moment of time when the laser field is the strongest, i.e. at $t=0$ or $t=\pi/\omega$. Detailed calculations, see e.g. \cite{KR07}, support this expectation. Without loss of generality we can presume that the pair is created at $t=0$; we also assume that the nucleus is located at the origin. This implies the initial condition $\mathbf{r}_\pm(0)=0$. Additionally, the initial momenta of the created pair should satisfy $\mathbf{P}_\pm(0)=\mathbf{p}_\pm$. These initial conditions are accounted for in Eqs.(\ref{r})-(\ref{PL}) by the chosen constants of integration.

We know that leptons of an \ee pair are predominantly created with momenta, which satisfy Eq.(\ref{soft}).
Let us examine a possibility of their collision in this case.
Eq.(\ref{soft}) implies that trajectories of leptons are characterized by one parameter, which can be taken as the $x$-component of the electron momentum $k\equiv k_-$. Other components of momenta are
$\boldsymbol{p}_\pm=(\mp k,0,q)$, where $q=(m^2+k^2-\bar{\eta}^2)/(2\bar{\eta})$ and $\eta_\pm=\bar{\eta}\equiv m/\sqrt{2}$.
Using these facts and Eqs.(\ref{PT}),(\ref{PL}) one finds that the longitudinal momenta and energies of both leptons are equal, $P_+(\varphi)=P_-(\varphi)$, $\varepsilon_+(\varphi)=\varepsilon_-(\varphi)$. 
Eq.(\ref{r}) ensures then that the $z$-components of the radius-vectors of leptons are same
for any $\varphi$, while the $y$-components remain zero, 
\begin{eqnarray}
\label{yz}
y_+(\varphi)=y_-(\varphi)=0~,\quad \quad ~	z_+(\varphi)=z_-(\varphi)~.
\end{eqnarray}
Thus, the only condition, which needs to be satisfied  to ensure that the collision takes place is given by the $x$-components of coordinates, $x_+(\varphi)=x_-(\varphi)$. Using Eqs.(\ref{r}),(\ref{PT}) one rewrites it as an equation on $\varphi$
\begin{equation}
\label{k}
	k=-\frac{1}{\varphi}\int_0^\varphi eA(\varphi') \,d\varphi'=m\xi\,\frac{1-\cos\varphi}{\varphi}
	~.
\end{equation} 
Here $k=k_-$,  $\varphi=\omega(t-z)$ with $t>0$ and $z<t$, which implies $\varphi>0$. 
Eq.(\ref{k}) shows that for $k=0$ an infinite number of collisions with $\varphi=2\pi n$, $n=1,2, \dots$ are possible; with the increase of $k$ the number of possible collisions decreases; collisions do not take place when  either $k > 0.725\, m\xi$, or $k<0$, where $0.725 = \max\big((1-\cos\varphi)/\varphi\big)$.  

We conclude that if momenta of leptons satisfy Eq.(\ref{soft}) 
then the leptons definitely collide provided $0\le k\le 0.725\,m\xi$. The total instant energy $\varepsilon(\varphi) =2\varepsilon_-(\varphi)$ of colliding leptons equals
\begin{eqnarray}
\label{etot}
\varepsilon(\varphi) 
=\sqrt{2}\,m \big[\,\xi^2\,\big(\,(1-\cos\varphi)/\varphi- \sin\varphi\big)^2+3/2\,\big]	~.
\end{eqnarray}
We take into account here that  $\varepsilon_-(\varphi)=Q_-( \varphi )+\bar{\eta}$, 
use Eq.(\ref{PL}), and express $k$ via $\varphi$ using Eq.(\ref{k}). We see that at the moment of collision the energy of the pair can be high, $\varepsilon(\varphi) \propto \xi^2 m\gg m$. The highest energy $\varepsilon(\varphi) = 2.21\, \xi^2 m$ is reached when 
$\varphi = 1.409\,\pi$ and $k= 0.290\, m\xi$. For $k=0$ collisions take place at much lower energy $\varepsilon(2\pi n)=3m/\sqrt{2}$.

Let us find the rate $W_\mathrm{c}$ with which the collision events take place. With this purpose we need to multiply the \ee creation probability 
by a factor $f$, which accounts for the fact that the pair would eventually collide, $dW_\mathrm{c}=f\,dW_{e^+e^-}$. Clearly, $f$ is proportional to the cross section $\tilde \sigma$ of a collision process considered.
At the moment of collision the impact parameter $b$ of the \ee pair is small, $b^2\sim \tilde \sigma$. One deduces from this that the necessary factor reads 
$	f=\tilde {\sigma} \delta^{(2)}
	(\mathbf{b})$, which can also be conveniently rewritten as
	 $f=\int_0^{\infty}  \!\tilde \sigma \,v\,
\delta \big(\,\mathbf{r}_{+}(\varphi)-\mathbf{r}_{-}(\varphi)\,\big)dt$.	
Here $v=|\mathbf{v}|=|\mathbf{v}_+-\mathbf{v}_-|$ is the collision velocity of the \ee pair, a two-dimensional vector $\mathbf{b}$, which is
orthogonal to the collision velocity, $\mathbf{v}\cdot \mathbf{b}=0$, represents an impact parameter for the \ee collision. The integration over the moment of time of collision $t$ is introduced in $f$ to simplify calculations.
Calculating $W_\mathrm{c}=\int f\,dW_{e^+e^-}$ using Eq.(\ref{dW}) one can integrate over
electron momenta $d^3p_-$ with the help of the delta-function from $f$, which gives the determinant $\det\big( \partial \boldsymbol{p}_-/ \partial \boldsymbol{r}_{-} \big)$. It is convenient to rewrite the integral over $dt$ in terms of integration over the electron momentum $dk_-$, which brings in the factor $\partial t/\partial k_-$. This determinant and derivative are calculated directly from the classical trajectory Eqs.(\ref{r})-(\ref{PL}). The final result reads
\begin{eqnarray}
	\label{fin}
	W_\mathrm{c}=B\int
	\! \exp \bigg(\!-\frac{ 2\sqrt{3}}{\F}\frac{\bar{\cal{A}}}{m^2 }\bigg)\,
	\frac{\sigma\, \varepsilon \,m\, w}{(t-z)^2\,v} 
	\,	\frac{d^3 p_{+}dk_{-}\!}{(2\pi)^6}~.
\end{eqnarray}
Here $t=t(\varphi)$ and $z=z(\varphi)$  are the moment of time and $z$-coordinate of \ee collision, $\varepsilon=\varepsilon(\varphi)$ is the instant energy of the \ee pair at the moment of collision, $\sigma$ is the cross section in the center of mass  reference frame. The transition from the cross section in the laboratory reference frame $\tilde \sigma$ to  $\sigma$ prompts the appearance of a modified velocity $w=[v^2-(\mathbf{v}_+ \times\mathbf{v}_-)^2]^{1/2}$, see e.g. \cite{LL02-87-sh}. 

The integration in Eq.(\ref{fin}) runs over all possible momenta that lead to a collision.  
This includes any configuration of momenta, which satisfy Eqs.(\ref{soft}),(\ref{k}). 
Moreover, using the classical trajectories Eq.(\ref{r})-(\ref{PL}) it it easy to verify that if electron and positron momenta are close to a configuration specified by these equations (which is very probable, as Eqs.(\ref{dW}) shows), then a collision can definitely take place provided the momenta of the \ee pair satisfy 
\begin{eqnarray}
	\label{ke}
\kappa_-= \kappa_+~,\quad \delta\eta_-=\delta\eta_+~. 
\end{eqnarray}
The factor $\bar{\cal{A}}$ in Eq.(\ref{fin}) equals $\cal{A}$ in Eq.(\ref{A}) evaluated for these particular momenta. This discussion ensures that there is no additional suppression in the integral over 
$d^3p_+dk_-$ in Eq.(\ref{fin}), compared to the integral over same four variables in Eq.(\ref{dW}). Both these  integrals are saturated in approximately same integration
volume. As a result, we can find an estimate for the ratio $R$ of the number of collisions of \ee pairs to the total number of created \ee pairs simply by comparing the integrands 
in  Eqs.(\ref{dW}),(\ref{fin}) for this 4D integration
\begin{eqnarray}
	\label{R}
	R = \frac{ W_\mathrm{c} }{ W_{ e^+e^- }}\approx
	\frac{ \sigma\, \varepsilon \,m\,w}{(t-z)^2\,v} \frac{1}{\F m^2}\approx
	\frac{ \sigma \, \omega \,m}{\xi}~.
\end{eqnarray}
The factor $\F m^2$ in the denominator of the middle expression here arises from 
an integration over $\kappa_-$ and $\eta_-$
in Eq.(\ref{dW}). The last equality in Eq.(\ref{R}) takes into
account that $t\sim z\sim \xi^2/\omega$, $t-z\sim 1/\omega$, as 
Eqs.(\ref{r})-(\ref{PT}) show, and that according to Eq.(\ref{etot}) $\varepsilon \sim \xi^2m$, 
which also implies $w\simeq v\simeq 1$. In Eq.(\ref{R}) $\sigma$  describes a  collision process considered, see examples in (\ref{qe})-(\ref{mumu}) below.

A simple estimation $R_0$ for a relative number of \ee collisions compares $\sigma$ with a distance $z$ to the point of collision,  giving $R_0 =  \sigma /z^2\approx \sigma
\omega^2/\xi^4$, which is much smaller than Eq.(\ref{R}) is predicting. Thus, Eq.(\ref{R}) shows that there exists a large enhancing factor
${\cal K}$, which makes collisions more probable 
\begin{eqnarray}
\label{K}
R \approx  {\cal K}\,R_0~,\quad\quad {\cal K}=\xi^4/\F\gg 1~.
\end{eqnarray}
An analysis of Eq.(\ref{R}) reveals two physical reasons, which combine to produce the enhancement. Eq.(\ref{dW}) ensures that momenta of the leptons created are necessarily close to a configuration given by Eq.(\ref{soft}), in which case a collision is almost imminent (since the necessary additional condition Eq.(\ref{k}) is satisfied very easily). There is also a relativistic factor, which arises due to stretching of lepton trajectories along the $z$-axis that makes collisions of electrons and positron even more probable. 

Collisions of \ee pairs bring interesting implications. First of all there are quasielastic collisions
\begin{eqnarray}
	\label{qe}
	e^+ + e^-\rightarrow 	e'^{\,+}  +  e'^{\,-}~. 
\end{eqnarray}
Any  collision process of this type takes only a small fraction of the period of laser oscillations. Therefore the conservation laws in a collision process are applied to the instant energy and instant momentum. A redistribution of these quantities in the reaction (\ref{qe}) may lead to a strong  increase of a lepton quasienergy  $ \bar {\varepsilon}_\pm $, which is defined as the energy averaged over the period of laser oscillation, $\bar {\varepsilon}_\pm=\int_0^{2\pi}\varepsilon_\pm(\varphi)\,d\varphi/(2\pi)$.
To illustrate this statement assume that the momenta of a pair of leptons satisfy Eq.(\ref{soft}). Then using kinematics arguments  one finds that the quasienergy of the lepton in the final state  can be as large as $\bar{\varepsilon}_\pm'=\varepsilon(\varphi)+\xi^2 k^2/(\sqrt{2}m)$. Here $\varepsilon(\varphi)$ is the instant energy of the initially created \ee pair. The term $\propto \xi^2 k^2$ in the quasienergy can become parametrically large provided the frequency of the laser field is sufficiently low $\omega \ll (E/E_c)^{3/2}\,m$. However, even if this condition is not satisfied, the increase of the quasienergy of the lepton can be quite sizable because the instant energy can be higher than the quasienergy.

There are also inelastic collisions. An important example is the creation of $\gamma$-quanta
\begin{eqnarray}
	\label{2g}
	e^+ + e^-\rightarrow 	2\,\gamma~. 
\end{eqnarray}
The energy and momentum of photons created in this reaction equal the instant energy and momentum of the \ee pair at the moment of collision, which can be large prompting the energy of $\gamma$-quanta to be large as well. 
Eq.(\ref{etot}) shows that the highest possible instant energy of the pair is $\varepsilon(\varphi)=\ 2.21 \xi^2 m$, which gives the upper limit for the energy of each photon $\varepsilon_{\gamma,\,\mathrm{max}}=1.10\, \xi^2 m$.

It is important that (\ref{2g}) provides the most probable way for production of high-energy photons by a laser and Coulomb fields. Indeed, the high-energy photons cannot be produced in the region of small distances. Their production there can be considered as an effective addition to the mass of leptons, $m\rightarrow m+\delta m$, which inevitably results in exponential decrease of the probability Eq.(\ref{W}). The high-energy $\gamma$-quanta also cannot be produced during propagation of a lepton in the laser field.  The adiabatic nature of this propagation ensures exponential suppression of this process. In contrast, the photon yield through reaction (\ref{2g}) is relatively large, as Eq.(\ref{R}) shows. 

Another interesting example is the creation of heavy particles, for example muons in the process
\begin{eqnarray}
\label{mumu}
	\e^+ + e^-\rightarrow \mu^+ + \mu^-.
\end{eqnarray}
To make this process possible the instant energy of the \ee pair in the center of mass reference frame should exceed the threshold of the \mm production. The latter is given by the minimum of the instant energy of muons, which equals $2m_\mu$. From Eqs.(\ref{k})-(\ref{etot}) one deduces that the center of mass instant energy of the \ee pair is
$\varepsilon_\mathrm{scm}(\varphi)=2(m^2+k^2)^{1/2}$, which is reduced to $2 k$ provided $k\gg m$. Thus the threshold condition for \mm production restricts $k$ from below $k\ge m_\mu$. At the same time Eqs.(\ref{dW}),(\ref{A}) restrict it from above, $k\lesssim m\,\xi\, \sqrt{\F}$. 
Combining both conditions, one concludes that the muon production via collisions of \ee pairs is probable provided $ m_\mu/m \lesssim \xi \sqrt \F$. This condition can be understood as a restriction on the laser frequency
\begin{eqnarray}
	\label{om}
\omega ~\lesssim~  (E/E_c)^{3/2}\,m^2/m_\mu\ll m~.
\end{eqnarray}  
Energy absorbed from the laser field when the \ee pair is initially created  is typically $\varepsilon \simeq \xi^2 m/\sqrt{2}$ \cite{KR07}. 
Meanwhile, the energy of particles in the final state of all phenomena considered in (\ref{qe})-(\ref{mumu}) can be significantly higher. Thus, collisions of leptons of the \ee pair result in additional absorption of energy from the field. 

This fact reminds us again that in the processes discussed the \ee pair acts as an antenna. We verified that it can efficiently accumulate high energy from a laser field and then transfer it to other particles. As a result, creation of muons and high-energy photons becomes possible, which is notable since their production through the conventional mechanism is suppressed exponentially. The \ee antenna removes this exponential suppression, making creation of high-energy particles much more probable.

I am thankful to D.J.Robinson for discussions. This work was supported by the Australian Research Council.


\end{document}